**Equity, Emissions and the Inflation Reduction Act**

**Authors:** Lucas Woodley[1,2], Chung Yi See[1], Daniel Palmer[3], and Ashley Nunes[1,2,4]

[1] Department of Economics
Harvard College
Cambridge, MA, 02138, USA

[2] Faculty of Arts and Sciences
Harvard University
Cambridge, MA, 02138, USA

[3] Groton School
Groton, MA, 01450, USA

[4] Center for Labor and a Just Economy
Harvard Law School
Cambridge, MA, 02138, USA

**Corresponding Author:** Ashley Nunes, anunes@fas.harvard.edu

**Abstract**

Preowned vehicles are - regardless of propulsion type – disproportionally purchased by low-income households, a group that has long been unable to purchase electric vehicles (EV). Yet, low-income households would disproportionally benefit from EV adoption given the operating costs savings offered by electrification. To help realize this benefit, provisions of the 2022 Inflation Reduction Act (IRA) offer preowned EV purchasing incentives. How effective might these efforts be? Leveraging data from the United States Census Bureau, the National Household Travel Survey, and the Greenhouse gases, Regulated Emissions, and Energy use in Technologies Model, we address this question. Our findings are fourfold. First, we demonstrate that although low-income households are more likely to benefit from preowned EV purchasing incentives offered by IRA, up to 8.4 million low-income households may be ineligible owing to heterogeneity in vehicle procurement pathways (i.e., where a vehicle is purchased from). Second, we show that program ineligibility risks preventing up to 113.9 million tons of $CO_2$e in lifecycle emissions reduction benefits from being realized. Third, we find that procurement pathways depend on vehicle price. More expensive preowned vehicles are purchased directly from commercial dealers, while less expensive preowned vehicles are purchased from private sellers. These procurement pathways matter because qualification for IRA's incentives necessitates purchasing solely from commercial dealers (versus private sellers). Fourth, we demonstrate that while incentives motivating preowned vehicle purchases from commercial dealers may be effective if the vehicle costs more than $6,000, this effectiveness diminishes at higher price points. The implications of our findings on decarbonization efforts and energy policy are discussed.

**Introduction**

In 2022, the United States government passed its most prominent piece of climate legislation to date: the Inflation Reduction Act (IRA). Among other provisions, IRA represents an investment in programs that incentivize clean energy and carbon management, reduce methane emissions, promote domestic supply chains, address environmental justice concerns, and most notably, encourage the adoption of electric vehicles (EVs) (1). This encouragement reflects longstanding environmental concerns surrounding the use of hydrocarbon powered vehicles (hereafter referred to as ICE vehicles).

Vehicles powered by hydrocarbons constitute one of the largest emitters of $CO_2$, a contribution that accelerates the greenhouse effect, causing global temperatures to rise (2). EVs, by contrast, offer a more favorable emissions profile that persists even after accounting for emissions associated with vehicle production, extraction, processing, transportation, and fuel distribution (2,3). However, widespread adoption of EVs is challenged, in part, by higher average up-front procurement prices (4). Higher prices are – from the vantage point of consumer adoption - particularly problematic for North American cities where the urban fabric and transportation infrastructure are tailored to the needs of conventional (and cheaper) fossil-fuel powered automobiles (5).

Recognizing this challenge, IRA offers – via amendments to the tax code (Section 30D) – up to $7,500 in incentives for consumers willing to purchase a new EV (6,7). However, a less discussed but arguably more important incentive offered by IRA is a credit dedicated solely towards the purchase of preowned EVs. Consumers who purchase electrified vehicles that have had a previous owner can – owing to a new provision in the tax code (Section 25E) – claim up to $4,000 or 30 percent of the sale price of the EV, whichever is lower, in savings (8).

These savings are particularly beneficial to low-income households who spend a disproportionately high percentage of their income on transportation-related expenses (9). Although EV adoption offers relief via more favorable operating costs (10-12), low-income households have long been priced out of EV procurement programs (13) due to new EVs' high up-front prices (14). Consequently, 25E offers an opportunity to distribute EV incentives more equitably by stimulating electrification demand in the preowned vehicle market, one that low-income households favor for vehicle purchases (15-17).

However, while the more equitable distribution of benefits features prominently in justifications for IRA (1), inclusion of low-income households in electrification efforts is timely for another reason. By virtue of being older, preowned vehicles have lower fuel economy. Ceteris paribus, this makes preowned vehicles more polluting than their newer counterparts given the interdependencies between fuel economy and tailpipe emissions (18). Consequently, IRA offers via 25E an opportunity to replace relatively polluting vehicles with less polluting ones, thereby increasing emissions reductions.

How likely is this prospect? To what extent may low-income households benefit from the 25E provision? Will some of these households be excluded? If so, how many? And what public policies would be most effective in maximizing uptake of 25E credits among these households? We leverage results from existing studies to address these questions (17). In doing so, we scrutinize the procurement patterns in low-income households, the provisions of 25E, and the trajectory of vehicle pricing in the preowned market to assess the potential effectiveness of 25E in stimulating EV demand.

We distinguish our efforts from previous effort in two ways. First, we do not scrutinize enthusiasm for and likelihood of EV adoption given differences in their aesthetic and performance profile relative to their ICE counterparts (19,20). We also do not emphasize household willingness to purchase an EV versus an ICE vehicle (21). While treatment of these issues is timely, should be debated, and warrant further scrutiny, it is not the focus of our efforts. Rather, we assess whether – solely from the vantage point of upfront vehicle procurement price – the structure of provisions in 25E can facilitate EV adoption among low-income households given the purchasing patterns observed among these households. This approach – which to our knowledge has not been previously executed - allows us to determine the effectiveness of programs like 25E.

In doing so, we clarify the three main requirements of 25E. First, the purchaser must have a taxable income of less than $150,000 for joint filers ($75,000 for single filers); second, the vehicle must be purchased from a commercial dealer[1]; and third, the vehicle must cost less than $25,000 (9)[2]. Given our focus on vehicle purchase price and seller among low-income households, defined as households that earn less than $40,000 annually[3], the first requirement is satisfied. We account for the second requirement by scrutinizing the procurement pathway of these households (i.e., whether their vehicles are purchased from private sellers or commercial dealers). Finally, we account for the third requirement by analyzing the long-run trajectory of used vehicle prices and reconciling this trajectory with the price cap specified in 25E.

We emphasize the timeliness of our efforts given IRA represents the first piece of legislation in the United States to explicitly incentive the adoption of secondhand vehicles. Although the durability of the 25E provision remains unclear – this owing to a recent change in the composition of the executive and legislative branch, - scrutinizing the potential of provisions like 25E represents, we argue, an effort to assess whether efforts that deliberately target a nuanced portion of the auto market may facilitate emissions reductions, and if so, to what extent.

---

[1] Section 30D(g)(8) defines the term "dealer" as a person licensed by a State, the District of Columbia, the Commonwealth of Puerto Rico, any other territory or possession of the United States, an Indian tribal government, or any Alaska Native Corporation (6). Private parties may leverage non-dealer pathways to realize the 25E credit. Doing so – we note – would violate the spirit of the 25E provision.

[2] 25E also specifies that the car be at least two years old. However, meeting this threshold is less onerous given that the average age of a used car at the point of purchase is 6.47 years (22). Compliance with the $25,000 price cap requirement is more challenging as average used vehicle prices are – year on year - more likely to exceed this threshold.

[3] This demographic – which constitutes 27 percent – of households across America, reflects the upper income limit of working low wage jobs, 40 hours of work a week, 50 weeks a year (23,24).

**Method**

Our approach consists of three parts. First, we estimate eligibility (expressed in relative terms) for 25E by scrutinizing the procurement patterns among low-income households. Second, we estimate – leveraging data from United States Census Bureau and the 2022 National Household Travel Survey (NHTS) – the absolute number of households who would be eligible for credit qualification. Third, we leverage the Greenhouse gases, Regulated Emissions, and Energy use in Technologies Model (GREET) to assess the emissions impact of EV adoption among the low-income households identified in Parts 1 and 2.

Part 1: We use publicly available survey data that scrutinizes the auto purchasing patterns among low-income households (17). Low-income households are defined – for the purposes of the survey – as being those that earn less than $40,000 annually. This demographic, which constitutes 27 percent of households across America, reflects the upper earnings limit of a dual income household where both earners work minimum wage jobs, 40 hours of work a week, 50 weeks a year (22,23).

In the survey, 1,018 of these households are probed about, 1) how they acquired vehicles, 2) what the purchase price of the vehicle was, and 3) whether the vehicle was purchased from a commercial dealer or private seller (20). We omit responses from households that could not recall the purchase price of their cars or who acquired their cars without payment (such as through work, inheritance, as a gift, etc.). This reduces the applicable sample size to 738 survey respondents. Leveraging this sample, we subsequently execute the following steps.

Respondents are assigned to one of 11 groups, each of which denotes a specific vehicle price bracket specified in the original survey (e.g., $500 - $999) (Table 1a and 1b, Step 1). Price ranges associated with each of these 11 brackets are subsequently adjusted upwards by $4,000 to reflect the full monetary value associated with the 25E provision[4] (Table 1a and 1b, Step 2). For example, the "$500 to $999" bracket now becomes $4,500 to $4,999 and the $1,000 to $1,499 bracket becomes $5,000 to $5,499. However, the number of respondents in the adjusted price bracket remains unchanged (Step 2). For example, in the original survey data, 1 household reports spending between $500 to $999 on a preowned vehicle that is acquired directly from a dealer. After becoming eligible for a $4,000 incentive, that household is – for the purposes of our analysis – assumed to spend between $4,500 to $4,999 on a preowned vehicle acquired directly from a dealer.

We subsequently revise the number of price brackets. This approach is necessary as not all 11 original price brackets have common upper and lower limits as the 11 adjusted price brackets. In assessing commonality, we observe that overlaps occur at $0, $5999/$6,000, $7999/$8,000, and $9,999/$10,000. Accounting for these overlaps creates 4 new price brackets: namely, (1) Less than $5,999, (2) $6,000 to $7,999, (3) $8,000 to $9,999, and (4) $10,000 or more (Table 1a and 1b, Step 3).

Having revised the number of price brackets (from 11 to 4), these brackets are subsequently populated with respondent data. This entails accounting for spending patterns before and after the introduction of 25E (Table 1a and 1b, Step 4). The following approach is executed. First, respondents in the original 11 price brackets (Step 1) are sorted into these 4 new brackets that do not account for 25E (Step 3). That is, data from Step 1 is leveraged to populate cells in Step 3. Subsequently, respondents in the 11 adjusted price brackets (Step 2) are sorted into these 4 new brackets that do account for 25E (Step 4). That is, data from Step 2 is leveraged to populate cells in Step 4. By scrutinizing changes in proportion of respondents in the 4 price brackets, we can assess changes in respondent spending patterns owing to 25E credit availability.

---

[4] Doing so – we recognize assumes perfect incidence: that is, that the totality of the $4,000 credit made available by the 25E provision is realized by the consumer (compared to more limited pass-through (35).

For example, in our original data set (which has 11 price brackets), 91 respondents reported spending between "$2,000 to $3,999" on their vehicle. Of that 91, 2 purchased their vehicle new directly from a commercial dealer, 56 preowned from a private seller, and 33 preowned from a commercial dealer. Owing to the availability of 25E, those 33 respondents would still be able to purchase a preowned vehicle at a dealer, this time spending between $6,000 to $7,999. However, the 2 respondents who purchased their vehicle 'new from a dealer' in the "$2,000 to $3,999" bracket remain in this bracket since they are ineligible for the $4,000 incentive owing to their purchasing method (i.e., new versus preowned).

We acknowledge that this approach assumes universal access to the $4,000 credit and ignores the potential limit of 30 percent of a vehicle's selling price. This suggests that a household purchasing a preowned vehicle for less than $13,333.33 will receive less than $4,000. However, such vehicles are exceedingly rare, as most preowned EVs cost more than $13,333.33. Thus, our data and conclusions do not meaningfully change when one relaxes this assumption.

Moreover, this approach allows us to estimate changes in the proportion of respondents choosing each purchase method in each new price bracket. This is done by multiplying the proportion of respondents in that price bracket with the proportion of respondents choosing each purchase method in that price bracket. Finally, we can compare the real proportion of respondents choosing each purchase method before and after the shift and analyze how the 25E provision affected a consumer's choice of purchase method.

Part 2: Here, we use data from the United States Census Bureau data and the 2022 NHTS to estimate the total number of households who would be eligible for credit qualification (Table 2). According to census data, the total number of households in the United States is 125,736,353. Households earning less than $40,000 annually constitute 26.7 percent of these households, or 33,571,606 (23,24). Leveraging these figures, we now estimate the proportion of low-income households that own a vehicle. We estimate that between 80.92 percent and 85.75 percent of low-income households own a vehicle, which corresponds to 27,166,144 to 28,787,652 households. Finally, using the survey data in Part 1, we estimate that 77.91 percent of respondents own a preowned vehicle. Among these 77.91 percent of respondents, 37.57 percent purchased their preowned car from a private seller and thus would be ineligible for financial relief from IRA owing to their procurement of vehicles from private sellers (versus commercial dealers). We estimate that this prohibits access to 25E for between 7,951,744 to 8,426,372 households in total.

Part 3: Finally, leveraging data from GREET, we assess the emissions impact of EV adoption among these low-income households (see Table 3 for the main assumptions and results) (36). By comparing the emissions profile of a preowned ICE vehicle to a preowned EV, we can calculate the lost lifecycle emissions benefit if a household were unable to access 25E and thus did not replace their preowned ICE vehicle with a preowned EV.

We begin by estimating the remaining lifecycle emissions for a preowned ICE vehicle, assuming that the ICE vehicle was purchased in 2014, has been owned for 10 years, and has 5 years left in its lifecycle.

These assumptions are consistent with procurement and utilization trends observed among low-income households today (see Table 3). We leverage existing data on emissions associated with fuel usage for ICEVs. In line with previous efforts (18,37,38), we assume for ICE vehicles an average emissions rate of 73 g$CO_2$e/MJ from fuel usage, 19 g$CO_2$e/MJ from fuel production, and a fuel economy of 34.6 miles per gallon (MPG). We also assume 8 metric tons of $CO_2$e from manufacturing emissions and an aggregate utilization of 179,200 miles travelled over 15 years. We estimate total per-mile emissions, considering per-mile vehicle manufacturing emissions, fuel usage and production emissions, fuel economy, and aggregate utilization using the following equation:

$$E_{PM} = \frac{(e_{vm} \times 1{,}000{,}000)}{au} + \left(\frac{1}{FE} \times (e_{fp} + e_{fu})\right)$$

where $E_{PM}$ = emissions per mile (g$CO_2$e/mi); $e_{vm}$ = vehicle manufacturing emissions (tons $CO_2$e); au = aggregate utilization (miles); FE = fuel economy (MPG); $e_{fp}$ = fuel production emissions (g$CO_2$e/MJ); and $e_{fu}$ = fuel usage emissions (g$CO_2$e/MJ).

We then calculate the remaining lifecycle emissions using the following equation:

$$E_{PV} = \frac{au_5}{1{,}000{,}000} \times E_{PM}$$

where $E_{PV}$ = emissions per vehicle (tons $CO_2$e); $au_5$ = aggregate utilization of the vehicle in its remaining 5 years, which we assume to be 50,200 miles; and $E_{PM}$ = emissions per mile (g$CO_2$e/mi). Given these assumptions, we find that the remaining lifecycle emissions of an ICE vehicle is 18.43 tons of $CO_2$e per vehicle.

We then estimate the remaining lifecycle emissions for a preowned EV. Assuming a household would likely purchase a preowned EV with equivalent or less mileage than their current preowned ICE vehicle, we estimate the remaining lifecycle emissions for a preowned EV purchased in 2014 that also has 5 years left in its lifecycle. In line with previous efforts (25,39,40), we assume 12.64 metric tons of $CO_2$e from manufacturing emissions (including NMC811 battery manufacturing emissions), a fuel economy of 105.8 MPG, and an aggregate utilization of 179,200 miles to be travelled in the EV's 15-year lifecycle. Depending on the carbon intensity of the electrical grid, we estimate remaining life cycle emissions using an average fuel production emissions rate of 82.61 g$CO_2$e/MJ (assuming less grid decarbonization) as well as of 23.81 g$CO_2$e/MJ (assuming a cleaner grid). Total per-mile emissions and remaining lifecycle emissions for the preowned EV are estimated using the same formulas leveraged for ICE calculations.

Given these assumptions, we find that the remaining lifecycle emissions of a preowned EV is 4.91 to 8.30 tons of $CO_2$e per vehicle depending on electric grid decarbonization rates. We calculate the difference between the remaining lifecycle emissions of a preowned ICE vehicle and a preowned EV to be 10.13 to 13.52 tons of $CO_2$e per vehicle.

We conclude by estimating the total lifecycle emissions benefits that cannot be realized due to the vehicle procurement pathway requirements of 25E. To do so, we combine our lower- and upper-bound estimates of the number of households who cannot access 25E with the range of per-vehicle emissions benefits associated with pre-owned vehicle electrification. Accounting for the between 7,951,744 and 8,426,372 households unable to access 25E calculated in Part 2, we find that between 80.55 to 113.92 million tons of $CO_2$e of cumulative lifecycle emissions reduction benefits go unrealized.

**Results and Discussion**

Our analysis offers four key findings.

First, we find that low-income households are more likely to benefit from incentive programs that deliberately incentivize preowned (versus new) vehicle purchases. 77.91 percent of these households report purchasing a previously owned vehicle, compared to 22.09 percent who report buying their vehicle new. Although this finding – which is consistent with previous research (14-16) – implies that nearly four-fifths of low-income households would benefit from the 25E provision, we caution that more nuance is warranted. Qualification for 25E necessitates not only that the vehicle be preowned but also that the preowned vehicle be purchased from a commercial dealer. This provision imposes a more onerous qualification standard, which highlights the tendency for many low-income households to purchase vehicles from private sellers. Of the 77.91 percent of households that report purchasing a preowned car, 62.43 percent report making these purchases directly from a commercial dealer (Fig 1a). This implies that where preowned vehicles are concerned, 37.57 percent of low-income households may be ineligible for financial relief from IRA owing to their procurement of vehicles from private sellers (versus commercial dealers). We estimate that such a restriction renders between 7,951,744 to 8,426,372 low-income households ineligible.

What are the emissions consequences of this ineligibility? Our second finding highlights the climate impact of pricing low-income communities out of the 25E credit. Accounting for the age of vehicles procured by these households and the overall longevity of these vehicles (25-28), outstanding emissions (i.e., emissions the vehicle will generate from the procurement point until retirement) are estimated to be 18.43 tons of $CO_2$ if the vehicle is an ICE and 8.3 tons of $CO_2e$ if the vehicle is an EV. This implies a lost lifecycle emissions benefit – on a per vehicle basis – of up to 10.13 tons of $CO_2e$. Cumulative lost lifecycle emissions benefits - derived by accounting for 7,951,744 to 8,426,372 households being unable to access 25E – yields up to 85.35 million tons of $CO_2e$ in lifecycle emissions reduction benefits that go unrealized. This figure rises to 113.92 million tons of $CO_2e$ should the Inflation Reduction Act facilitate significant reductions in the carbon intensity of the electrical grid. This is because doing so would raise the per-vehicle emissions benefit of EVs to a maximum of 13.52 tons of $CO_2e$. Cumulative lost lifecycle emissions benefits can rise even further should technological advances also facilitate improvements in the weight profile of EVs, which improve fuel economy and, consequently, the emissions profile of the vehicle (27).

Ineligibility for 25E among low-income households reflects heterogeneity in the preferred purchasing pathway of preowned vehicles. What explains this heterogeneity?

Our third finding is that vehicle price influences from whom low-income households purchase preowned vehicles. Specifically, we find that the inclination to rely on commercial dealers (versus private sellers) for preowned vehicle purchases increases as vehicle price increases. When a preowned vehicle costs less than $5,999, 30.92 percent of households purchase these vehicles from commercial dealers. However, 67.50 percent of low-income households rely on commercial dealers when the vehicle costs between $6,000 and $7,999. A further increase in preowned vehicle price to between $8,000 to $9,999 yields the highest number of households (78.78 percent) willing to purchase these vehicles from commercial dealers (Fig. 1b).

Although this trend implies that additional increases in preowned vehicle price should increase the number of households willing to purchase these vehicles from commercial dealers, our results suggest otherwise. Specifically, we find that when the price of preowned vehicles exceeds $10,000, only 55.75 percent of consumers purchase vehicles from commercial dealers. What motivates this change? We find that a decrease in preowned vehicle purchases from commercial dealers is accompanied by a corresponding increase in the proportion of respondents who buy their vehicles new.

Consequently, our fourth finding is that incentive programs like 25E can motivate vehicle purchases from commercial dealers if the vehicle costs more than $6,000, but the effectiveness of such programs diminishes at higher price points. Assuming the totality of the $4,000 credit offered by 25E can be applied towards the purchase of a preowned vehicle (i.e., perfect incidence) and the preowned vehicle costs between $6,000 to $7,999, the credit invites an 18.77 percentage point increase in the proportion of respondents purchasing preowned cars from a commercial dealer (from 43.54 percent to 62.31 percent) (Fig. 1c). However, this increase slows to 14.03 percentage points when the vehicle costs between $8,000 to $9,999, and to 9.19 percentage points when the vehicle costs more than $10,000.

What explains the 25E provision's diminishing effect at higher price points? Absent the credit, an increase in new vehicle price from between $8,000 and $9,999 to above $10,000 invites a 21.41 percentage point increase in new vehicle purchases, and a 26.29 percentage point decrease in preowned vehicle purchases (Fig. 1c). When the $4,000 credit is made available, the percentage point increase in new vehicle purchases rises to 22.40, and the percentage point decrease in preowned vehicle purchases rises to 32.42. The minimal increase in purchase propensity of new vehicles suggests that the credit does not – at higher price points – induce a meaningful change in the proportion of consumers willing to purchase preowned (compared to new) vehicles. A possible explanation for this phenomenon might be asymmetric information in the preowned vehicle market – uncertainty surrounding the quality of the preowned vehicles (relative to new ones) may disincentivize their purchase (29). Alternatively, given the choice between purchasing a preowned versus new good, consumers typically gravitate towards the latter despite demonstratable aesthetic and performance similarities between these goods (30).

These results underscore misalignment between low-income households' vehicle procurement pathways and requirements of programs such as 25E. Absent policy changes expanding eligibility to private sellers, our model estimates that nearly 40 percent of low-income households will be unable to qualify for the preowned EV tax credit. Households purchasing vehicles that cost less than $5,999 will likely face the largest disparities, as nearly 70 percent of these households do not currently rely upon commercial dealers for auto purchases. Although the existing provisions of 25E could motivate households to change where they purchase their vehicles from (i.e., commercial dealers rather than private sellers), the plausibility of this outcome is unclear. This lack of clarity – which highlights the need for subsequent work - reflects longstanding legislative emphasis on solely motivating the purchase of vehicles (albeit with varying degrees of success rather than considering where those vehicles are procured from) (31).

**Limitations and Conclusion**

Our work leverages assumptions that reflect the constantly improving value proposition of EVs relative to ICEs. Notably, we assume easy access to recharging infrastructure, equivalency (or acceptable) energy replenishment rates for EVs compared to ICEs, and perfect incidence regarding incentive realization. However, we recognize the challenges in bringing these assumptions to fruition.

For example, while IRA allocates $7.5 billion towards building new recharging stations, an investment that disproportionately favors low-income households (given the lack of dedicated parking these households have), the true cumulative investment required to ameliorate public range anxiety concerns is estimated to be between $31 and $55 billion (32). Absent this investment, EV uptake may be impeded regardless of whether programs like 25E are available. A similar dilemma persists regarding energy replenishment rates. Whereas these rates are poised to improve as auto makers advance battery technology charging station architecture and charging methodology, whether these rates can ultimately match those offered by ICE equivalents remains uncertain. The lack of parity in energy replenishment rates between preowned EVs and preowned ICEs may disincentivize uptake given the high opportunity costs associated with EV recharging, thereby deemphasizing the potential of programs like 25E. We recognize that weakening one (or both) these assumptions make preowned EV uptake by low-income households more challenging.

Furthermore, we caution that when scrutinizing procurement patterns and pathways in the auto market, we assume perfect incidence of 25E for low-income households. However, this assumption may be challenged by the relationship between new and preowned vehicle markets: the quantity of preowned vehicles that can be purchased at any given time is a direct function of how many new vehicles were purchased in a previous period. This implies that the short run supply of preowned EVs is fixed (i.e., perfectly inelastic) at the initial time of a policy's enactment. Thus, to the extent that incentives such as 25E successfully increase demand for preowned EVs, inelastic supply suggests that preowned EV prices will simply increase by an equivalent amount (i.e., $4,000) (33,34). Consequently, low-income households are unlikely to realize an initial increase in purchasing power for preowned EVs, resulting in relatively less vehicle turnover and fewer emissions benefits. Even if preowned EV prices were held constant, we acknowledge that it is unlikely that the purchaser will spend 100 percent of the 25E incentive on a more expensive used vehicle. Put simply, assumptions regarding the distribution of the incentive's benefits between secondhand EVs' sellers and buyers (i.e., perfect incidence) may result in overestimation of the expected near-term impact of 25E among low-income households.

The efficacy of 25E may be further diminished over time given the rising long run trajectory of used vehicle prices. These prices have – in inflation adjusted dollars – increased from $21,493 in 2019 to $25,891 in 2021 to $26,700 in 2023, a finding that would – ceteris paribus - further restrict access to the 25E credit, given the $25,000 price cap required for credit qualification. We also acknowledge that IRA credits phase out in 2032 and have not considered the availability and price of used vehicles at the program's expiration.

Furthermore, our approach focuses on the potential for 25E to achieve its intended goal: to effectively spur EV demand among low-income households. It is possible that low-income households' purchasing decisions are broadly insensitive to the credit due to a strong preference for ICEs over EVs, regardless of each vehicle's features and cost. Conversely, if existing used EV demand vastly exceeds supply irrespective of available incentives, 25E is unlikely to produce significant emissions benefits or improve the equity of EV ownership. Nevertheless, to the extent that used EV adoption can be spurred by purchase incentives, our results highlight the promise programs like 25E offer. Specifically, we find that such an incentive program that deliberately targets the procurement of preowned vehicles disproportionally benefit low-income households.

Finally, we note that our analysis leverages a data set that stratifies purchasing patterns among low-income households (i.e., those earning less than $40,000 annually). This diminishes the number of data points in each strata which may raise questions about the generalizability of our results. Although the smaller sample may constrain the breadth of inferences, it nevertheless offers valuable insights into the purchasing behavior of low-income households. Moreover, we note that the dataset has been used in previous research (17) and is a representative sample that reflects auto purchasing patterns among low-income households.

Collectively, our findings have important implications for clean energy policies as policymakers aim to craft targeted interventions that promote equitable EV adoption. Policies that expand access to preowned EVs for low-income households can contribute significantly to emission reductions by replacing older, more polluting vehicles in these communities. Such efforts may be further supported by complementary energy policies that increase charging station accessibility or subsidize electricity costs for EV charging among low-income households.(18,35). However, as our findings illustrate, maximizing the emissions reductions potential of programs such as 25E necessitates, a) considering the procurement pathways used by these households, and b) tailoring EV adoption programs to accommodate these pathways. Absent doing so, inequities in EV adoption are likely to persist.

**References**

1. Bistline, John, et al. "Emissions and Energy Impacts of the Inflation Reduction Act." *Science*, 2023, https://doi.org/10.1126/science.adg3781. Accessed 6 Jan. 2024.

2. Wolfram, Paul, et al. "Pricing of indirect emissions accelerates low-carbon transition of US Light Vehicle Sector." *Nature Communications*, 2021, https://doi.org/10.21203/rs.3.rs-334331/v1.

3. Koengkan, Matheus, et al. "The Impact of Battery-Electric Vehicles on Energy Consumption: A Macroeconomic Evidence from 29 European Countries." *World Electric Vehicle Journal*, vol. 13, no. 2, p. 36, https://doi.org/10.3390/wevj13020036. Accessed 6 Jan. 2024.

4. Editors, KBB. "Evaluating EV Facts over EV Hype." *Kelley Blue Book*, 16 March 2021, https://www.kbb.com/car-advice/ev-facts-over-ev-hype/. Accessed 5 January 2024.

5. Janushewski, Aaron Gordon. Auto-Centric Dependency: How Transportation Affected North American Cities. 2014. Carleton University, Master's Thesis. https://repository.library.carleton.ca/concern/etds/zp38wd26v?locale=en

6. "Text - H.R.5376 - 117th Congress (2021-2022): Inflation Reduction Act of 2022." *Congress.gov*, https://www.congress.gov/bill/117th-congress/house-bill/5376/text. Accessed 5 January 2024.

7. Trost, Jenna, and Dunn, Jennifer. "Assessing the Feasibility of the Inflation Reduction Act's EV Critical Mineral Targets." *Nature Sustainability*, vol. 6, no. 6, 2023, pp. 639-643, https://doi.org/10.1038/s41893-023-01079-8. Accessed 25 Jul. 2023.

8. "Used Clean Vehicle Credit | Internal Revenue Service." *IRS*, 28 November 2023, https://www.irs.gov/credits-deductions/used-clean-vehicle-credit. Accessed 5 January 2024.

9. "Commuting Expenses: Disparity for the Working Poor." *BTS*, 9 November 2017, https://www.bts.gov/sites/bts.dot.gov/files/legacy/publications/special_reports_and_issue_briefs/issue_briefs/number_01/pdf/entire.pdf. Accessed 19 January 2024.

10. Bauer, Gordon, and Nic Lutsey. "When might lower-income drivers benefit from electric vehicles? Quantifying the economic equity implications of electric vehicle adoption." *International Council on Clean Transportation*, 1 February 2021, https://theicct.org/wp-content/uploads/2021/06/EV-equity-feb2021.pdf. Accessed 19 January 2024.

11. Borlaug, Brennan, et al. "Levelized Cost of Charging Electric Vehicles in the United States." *Joule*, vol. 4, no. 7, 2020, pp. 1470-1485, https://doi.org/10.1016/j.joule.2020.05.013. Accessed 15 Jan. 2024.

12. Lanz, Lukas, et al. "Comparing the Levelized Cost of Electric Vehicle Charging Options in Europe." *Nature Communications*, vol. 13, no. 1, 2022, pp. 1-13, https://doi.org/10.1038/s41467-022-32835-7. Accessed 15 Jan. 2024.

13. Guo, Shuocheng, and Eleftheria Kontou. "Disparities and equity issues in electric vehicles rebate allocation." *Energy Policy*, vol. 154, July 2021, p. 112291, https://doi.org/10.1016/j.enpol.2021.112291.

14. "The Plug-in Electric Vehicle Tax Credit - Congress." *Congressional Research Service*, 2019, https://crsreports.congress.gov/product/pdf/IF/IF11017.

15. Pierce, Gregory, and Rachel Connolly. "Disparities in the "Who" and "Where" of the Vehicle Purchase Decision-making Process for Lower-income Households." *Travel Behaviour and Society*, vol. 31, 2023, pp. 363-373, https://doi.org/10.1016/j.tbs.2023.02.003. Accessed 15 Jan. 2024.

16. Wessel, Ryan J. "Policing the Poor: The Impact of Vehicle Emissions Inspection Programs across Income." *Transportation Research Part D: Transport and Environment*, vol. 78, 2020, p. 102207, https://doi.org/10.1016/j.trd.2019.102207. Accessed 6 Jan. 2024.

17. Klein, Nicholas J., et al. "In the Driver's Seat: Pathways to Automobile Ownership for Lower-income Households in the United States." *Transportation Research Interdisciplinary Perspectives*, vol. 18, 2023, p. 100787, https://doi.org/10.1016/j.trip.2023.100787. Accessed 15 Jan. 2024.

18. "Insights into Future Mobility." *Mobility of the Future*, MIT Energy Initiative, Nov. 2019, energy.mit.edu/wp-content/uploads/2019/11/Insights-into-Future-Mobility.pdf.

19. Hardman, Scott, and Gil Tal. "Understanding Discontinuance among California's Electric Vehicle Owners." *Nature Energy*, vol. 6, no. 5, 2021, pp. 538-545, https://doi.org/10.1038/s41560-021-00814-9. Accessed 15 Feb. 2024.

20. Hardman, Scott, et al. "A Review of Consumer Preferences of and Interactions with Electric Vehicle Charging Infrastructure." *Transportation Research Part D: Transport and Environment*, vol. 62, 2018, pp. 508-523, https://doi.org/10.1016/j.trd.2018.04.002. Accessed 15 Feb. 2024.

21. Linn, Joshua. "Is There a Trade-Off Between Equity and Effectiveness for Electric Vehicle Subsidies?" *Resources for the Future Working Paper*, 7 January 2022, https://media.rff.org/documents/WP_22-7.pdf. Accessed 18 February 2024.

22. Papandrea, Dawn. "The Average Age of a Used Car at Purchase Is 6.47, With Wide Variation Among States and Metros." *ValuePenguin*, 29 August 2022, https://www.valuepenguin.com/used-car-ages-study. Accessed 14 February 2024.

23. "Household Income: HINC-06." *U.S. Census Bureau*, 9 August 2023, https://www.census.gov/data/tables/time-series/demo/income-poverty/cps-hinc/hinc-06.html. Accessed 16 January 2024.

24. "Brief State Minimum Wages." *National Conference of State Legislatures*, 1 January 2024, https://www.ncsl.org/labor-and-employment/state-minimum-wages. Accessed 13 February 2024.

25. Elgowainy, Amgad, et al. "Current and future United States light-duty vehicle pathways: cradle-to-grave lifecycle greenhouse gas emissions and economic assessment." *Environmental science & technology* 52.4 (2018): 2392-2399

26. Nunes, Ashley, et al. "Re-thinking procurement incentives for electric vehicles to achieve net-zero emissions." *Nature Sustainability*, vol. 5, no. 6, 2022, pp. 527–532, https://doi.org/10.1038/s41893-022-00862-3

27. "Insights into Future Mobility." *Mobility of the Future*, MIT Energy Initiative, Nov. 2019, energy.mit.edu/wp-content/uploads/2019/11/Insights-into-Future-Mobility.pdf.

28. "U.S. households are holding on to their vehicles longer." *U.S. Energy Information Administration (EIA)*, 21 August 2018, https://www.eia.gov/todayinenergy/detail.php?id=36914. Accessed 27 April 2024.

29. Akerlof, George A. "The market for 'Lemons': Quality Uncertainty and the market mechanism." *The Quarterly Journal of Economics*, vol. 84, no. 3, Aug. 1970, p. 488, https://doi.org/10.2307/1879431.

30. Fernando, Angeline. "Comparison of perceived acquisition value sought by online second-hand and new goods shoppers." *European Journal of Marketing*, vol. 52, no. I, 2018. *Research Gate*, https://www.researchgate.net/publication/324475710_Comparison_of_perceived_acquisition_value_sought_by_online_second-hand_and_new_goods_shoppers. Accessed 14 February 2024.

31. Seo, Boyoung, and Matthew H. Shapiro. "Minimizing Fleet Emissions through Optimal EV Subsidy Design and Vehicle Replacement." (2019). Working Paper.

32. Wood, Eric, et al. *The 2030 National Charging Network*, National Renewable Energy Laboratory, June 2023, www.nrel.gov/docs/fy23osti/85654.pdf. (WAS 33)

33. Woodley, Lucas. "Estimating the Impact of the Inflation Reduction Act on Used and New Vehicle Markets: Combining Theory and Fieldwork." Society for Judgement and Decision-Making Conference, November 2023. Conference Presentation.

34. Woodley, Lucas. "*Estimating the Impact of the Inflation Reduction Act on Used and New Vehicle Markets: Combining Theory and Fieldwork.*" 2023. Harvard U, Undergraduate honors thesis.

35. Pierce, Gregory, et al. *Designing Light-Duty Vehicle Incentives for Low- and Moderate-Income Households*, UCLA Luskin Centre for Innovation, 12 Mar. 2019, innovation.luskin.ucla.edu/wp-content/uploads/2019/06/Designing_Light-Duty_Vehicle_Incentives_for_Low-and_Moderate_Income_Households.pdf.

36. "Argonne GREET Model." *Argonne National Laboratory*, Argonne National Laboratory, greet.es.anl.gov/.

37. Barwick, Panle Jia, et al. "Pass-through of Electric Vehicle Subsidies: A Global Analysis." *AEA Papers and Proceedings*, vol. 113, 2023, pp. 323-328. *American Economic Association*, https://www.aeaweb.org/articles?id=10.1257/pandp.20231064.

38. Woodley, Lucas, et al. "Targeted electric vehicle procurement incentives facilitate efficient abatement cost outcomes." Sustainable Cities and Society, vol. 96, Sep. 2023, p. 104627, https://doi.org/10.1016/j.scs.2023.104627.

39. Bieker, Georg, et al. "More bang for the buck: a comparison of the life-cycle greenhouse gas emission benefits and incentives of plug-in hybrid and battery electric vehicles in Germany.", The International Council on Clean Transportation, Mar. 2022, theicct.org/wp-content/uploads/2022/03/ghg-benefits-incentives-ev-mar22-2-1.pdf.

40. "U.S. Energy-Related Carbon Dioxide Emissions, 2022." U.S. Energy Information Administration, 29 Nov. 2023, www.eia.gov/environment/emissions/carbon/.

41. "National Household Travel Survey." *NHTS*, U.S. Department of Transportation, 2009, nhts.ornl.gov/download.shtml.

42. *2014 Fuel Economy Guide*, U.S. Department of Energy, 2014, www.fueleconomy.gov/feg/pdfs/guides/feg2014.pdf.

**Acknowledgments**

L.W. was supported by a National Science Foundation Graduate Research Fellowship under grant 2140743. A.N. was partially supported by the Breakthrough Institute. We thank Nickolas Klein and Michael Smart for the assistance in securing relevant data sets.

| | 11 Original Price Brackets (before 25E) | Less than $500 | $500 to $999 | $1,000 to $1,499 | $1,500 to $1,999 | $2,000 to $3,999 | $4,000 to $5,999 | | |
|---|---|---|---|---|---|---|---|---|---|
| Step 1 | Vehicle acquired new from a commercial dealer | 0 | 0 | 1 | 0 | 2 | 1 | | |
| | Vehicle acquired preowned from a private seller | 15 | 18 | 30 | 18 | 56 | 40 | | |
| | Vehicle acquired preowned from a commercial dealer (before 25E) | 0 | 1 | 3 | 11 | 33 | 33 | | |
| | 11 Adjusted Price Brackets (after 25E) | Less than $4,500 | | | | | $4,500 to $4,999 | $5,000 to $5,499 | $5,500 to $5,999 |
| Step 2 | Vehicle acquired preowned from a commercial dealer | 0 | | | | | 1 | 3 | 11 |
| | 4 Price Brackets (before 25E) | Less than $5,999 | | | | | | | |
| Step 3 | Vehicle acquired new from a commercial dealer | 4 (1.53%) | | | | | | | |
| | Vehicle acquired preowned from a private seller | 177 (67.56%) | | | | | | | |
| | Vehicle acquired preowned from a commercial dealer | *81 (30.92%)* | | | | | | | |
| | 4 Price Brackets (after 25E) | Less than $5,999 | | | | | | | |
| Step 4 | Vehicle acquired new from a commercial dealer | 4 (2.04%) | | | | | | | |
| | Vehicle acquired preowned from a private seller | 177 (90.31%) | | | | | | | |
| | Vehicle acquired preowned from a commercial dealer | *15 (7.65%)* | | | | | | | |

Table 1a

| | 11 Original Price Brackets (before 25E) | $6,000 to $7,999 | $8,000 to $9,999 | $10,000 to $14,999 | | $15,000 to $19,999 | | $20,000 or more |
|---|---|---|---|---|---|---|---|---|
| Step 1 | Vehicle acquired new from a commercial dealer | 1 | 0 | 13 | | 39 | | 106 |
| | Vehicle acquired preowned from a private seller | 12 | 8 | 9 | | 8 | | 2 |
| | Vehicle acquired preowned from a commercial dealer (before 25E) | 27 | 28 | 95 | | 80 | | 48 |
| | 11 Adjusted Price Brackets (after 25E) | $6,000 to $7,999 | $8,000 to $9,999 | $10,000 to $11,999 | $12,000 to $13,999 | $14,000 to $18,999 | $19,000 to $23,999 | $24,000 or more |
| Step 2 | Vehicle acquired preowned from a commercial dealer | 33 | 33 | 27 | 28 | 95 | 80 | 48 |
| | 4 Price Brackets (before 25E) | $6,000 to $7,999 | $8,000 to $9,999 | $10,000 or more | | | | |
| Step 3 | Vehicle acquired new from a commercial dealer | 1 (2.50%) | 0 (0.00%) | 158 (39.50%) | | | | |
| | Vehicle acquired preowned from a private seller | 12 (30.00%) | 8 (22.22%) | 19 (4.75%) | | | | |
| | Vehicle acquired preowned from a commercial dealer | *27 (67.50%)* | *28 (77.78%)* | *223 (55.75%)* | | | | |
| | 4 Price Brackets (after 25E) | $6,000 to $7,999 | $8,000 to $9,999 | $10,000 or more | | | | |
| Step 4 | Vehicle acquired new from a commercial dealer | 1 (2.17%) | 0 (0.00%) | 158 (34.73%) | | | | |
| | Vehicle acquired preowned from a private seller | 12 (26.09%) | 8 (19.51%) | 19 (4.18%) | | | | |
| | Vehicle acquired preowned from a commercial dealer | *33 (71.74%)* | *33 (80.49%)* | *278 (61.10%)* | | | | |

Table 1b

| Total Number of Households | 125,736,353 | |
|---|---|---|
| Total Number of Low-Income Households Earning under $40,000 annually | 33,571,606 | |
| | **Lower-Bound** | **Upper-Bound** |
| Total Number of Low-Income Households (<$40,000) that Own a Vehicle | 27,166,144 | 28,787,652 |
| Total Number of Low-Income Households (<$40,000) that Purchased a Preowned Vehicle from a Private Seller | 7,951,744 | 8,426,372 |

Table 2: Census bureau estimates of low income household distribution

| | | ICE | EV | |
|---|---|---|---|---|
| **Assumed Numbers** | Aggregate utilization (mi) | | 179,200[a] | |
| | Fuel economy (MPG) | 34.60[b] | 105.80[b] | |
| | Fuel usage emissions (gCO$_2$e/MJ) | 73[c] | 0 | |
| | Manufacturing emissions (tons CO$_2$e) | 8[c] | 12.64[c] | |
| | Fuel production emissions (gCO$_2$e/MJ) | 19[c] | **50 % cleaner grid** | **90 % cleaner grid** |
| | | | 82.61[d] | 23.81[d] |
| **Per-Mile Emissions Breakdown (gCO$_2$e/mi)** | Fuel production per-mile emissions | 66.60 | 94.70 | 27.30 |
| | Fuel usage per-mile emissions | 256 | 0 | 0 |
| | Vehicle manufacturing per-mile emissions | 44.60 | 70.50 | 70.50 |
| **Per Vehicle Total Per-Mile Emissions (gCO$_2$e/mi)** | | 367.20 | 165.30 | 97.80 |
| **Per Vehicle Total Lifecycle Emissions (179,200 miles, 15 years of ownership[a]) (tons CO$_2$e/vehicle)** | | 65.80 | 29.62 | 17.53 |
| **Per Vehicle Total Emissions for First 10 Years (129,000 miles travelled, after 10 years of ownership[a]) (tons CO$_2$e/vehicle)** | | 47.37 | 21.32 | 12.62 |
| **Per Vehicle Total Emissions for Remaining 5 Years (50,200 miles remaining, after 10 years of ownership[a]) (tons CO$_2$e/vehicle)** | | 18.43 | 8.30 | 4.91 |
| **Per Vehicle Remaining Lost Lifecycle Emissions Benefit (tons CO$_2$e/vehicle)** | | | 10.13 | 13.52 |
| **Cumulative Lost Lifecycle Emissions Benefit (Lower-Bound) (million tons CO$_2$e)** | | | 80.55 | 107.5 |
| **Cumulative Lost Lifecycle Emissions Benefit (Upper-Bound) (million tons CO$_2$e)** | | | 85.35 | 113.92 |

Table 3: Overview of model assumptions and carbon intensity estimates

[a] The average vehicle travels 179,200 miles over a 15-year lifespan, travelling fewer miles in each year (41).

[b] For vehicles purchased in 2014, the average fuel economy for an ICE vehicle is 34.60 MPG and 105.80 MPG for an EV (42).

[c] Fuel production, fuel usage and vehicle manufacturing emissions for ICE vehicles are estimated to be 19 g$CO_2$e/MJ, 73 g$CO_2$e/MJ and 8 tons $CO_2$e respectively. Manufacturing emissions for EVs are estimated to be 12.64 tons $CO_2$e (27).

[d] Leveraging the GREET model in combination with a model developed in previous works, assuming 50 percent grid decarbonization by 2030, fuel production emissions for an EV is estimated to be 82.61 g$CO_2$e/MJ. Assuming 90 percent grid decarbonization by 2030, fuel production emissions for an EV is estimated to be 23.81 g$CO_2$e/MJ (27,36,38).

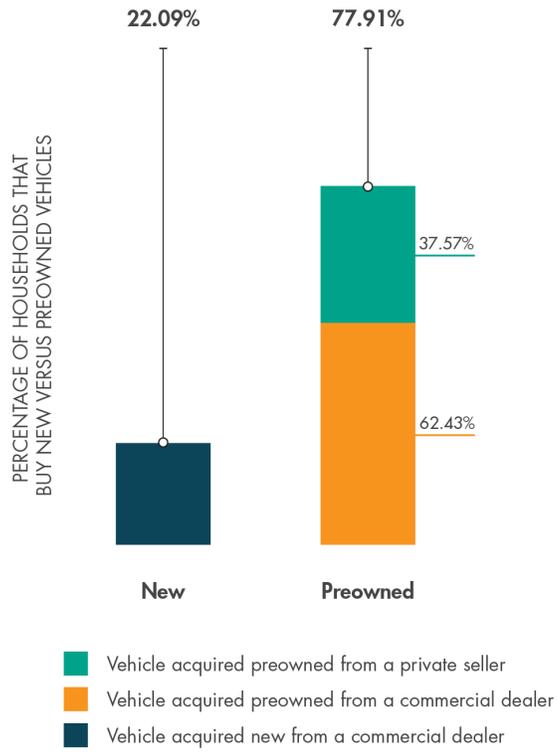

Fig 1a: Distribution of Respondents by Vehicle Purchase Method

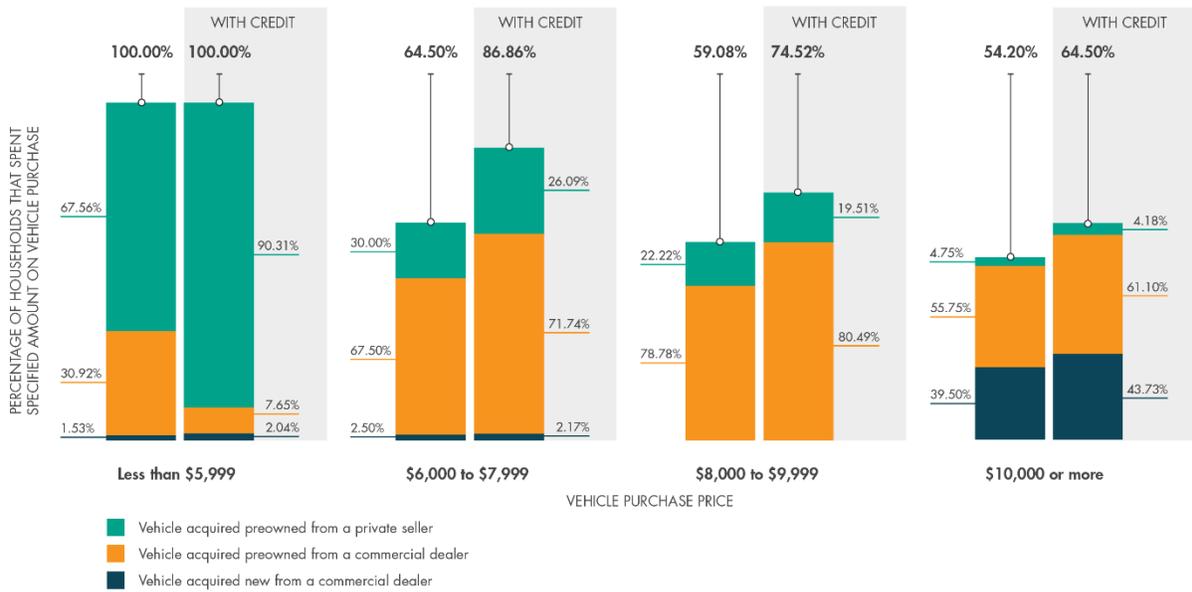

Fig 1b: Percentage of All Respondents who purchased a vehicle within specific price ranges

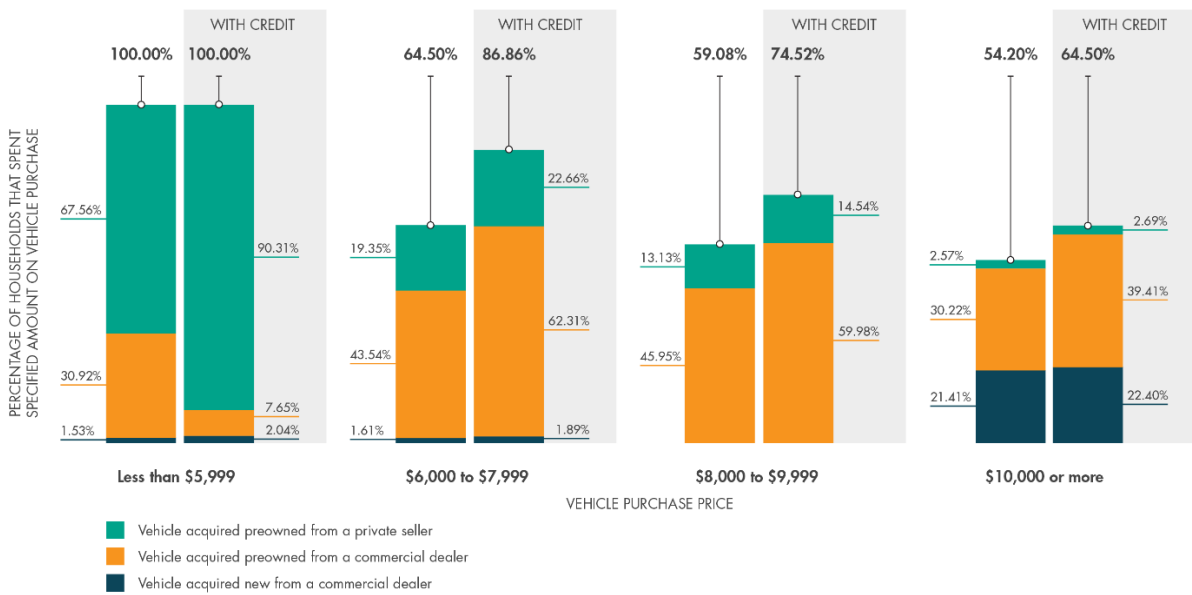

Fig 1c: Real Percentage of Respondents who purchased a vehicle within specific price ranges, stratified by procurement pathway and credit availability.

23